\documentclass[preprint2]{aastex}

\shorttitle{The linearity of the CMR of Virgo}
\shortauthors{Smith Castelli et al.}

\begin{document}

\title{About the linearity of the color--magnitude relation of early-type galaxies in the Virgo cluster}
\author{Anal\'ia V. Smith Castelli\altaffilmark{1}} 
\affil{Instituto de Astrof\'isica de La Plata (CCT-La Plata, CONICET-UNLP), Paseo del Bosque s/n, B1900FWA, La Plata, Argentina}
\email{asmith@fcaglp.unlp.edu.ar}
\author{N\'elida M. Gonz\'alez and Favio R. Faifer\altaffilmark{2}}
\affil{Facultad de Ciencias Astron\'omicas y Geof\'isicas, Universidad Nacional de La Plata, Paseo del Bosque s/n, La Plata, B1900FWA, Argentina}
\email{ngonzalez@fcaglp.unlp.edu.ar,favio@fcaglp.unlp.edu.ar}
\author{Juan Carlos Forte}
\affil{CONICET - Planetario de la Ciudad de Buenos Aires ``Galileo Galilei'', Av. Sarmiento y B. Rold\'an, Ciudad Aut\'onoma de Buenos Aires, Argentina}
\email{forte@fcaglp.unlp.edu.ar}
 
\altaffiltext{1}{Facultad de Ciencias Astron\'omicas y Geof\'isicas, Universidad Nacional de La Plata, Paseo del Bosque s/n, La Plata, B1900FWA, Argentina}
\altaffiltext{2}{Instituto de Astrof\'isica de La Plata (CCT-La Plata, CONICET-UNLP), Paseo del Bosque s/n, B1900FWA, La Plata, Argentina}

\begin{abstract}
We revisit the color--magnitude relation (CMR) of the Virgo cluster 
early-type galaxies in order to explore its alleged non-linearity. 
To this aim, we reanalyze the relation already published 
from data obtained within the {\it ACS Virgo Cluster Survey} of the Hubble 
Space Telescope, and perform our own photometry and analysis of the images 
of the 100 early-type galaxies observed as part of this survey. In addition,
we compare our results with those reported in the literature from data of 
the {\it Sloan Digital Sky Survey}. We have found that when the brightest 
galaxies and untypical systems are excluded from the sample, a linear relation 
arises in agreement with what is observed in other groups and clusters. The 
central regions of the brightest galaxies also follow this 
relation. In addition, we notice that Virgo contains at least four compact 
elliptical galaxies besides the well known object VCC\,1297 (NGC\,4486B). 
Their locations in the $\langle\mu_{\rm eff}\rangle$--luminosity diagram 
define a different trend to that followed by {\it normal} early-type dwarf 
galaxies, setting an upper limit in effective surface brightness and a lower 
limit in effective radius for their luminosities. Based on the 
distribution of different galaxy sub-samples in the color--magnitude and 
$\langle\mu_{\rm eff}\rangle$--luminosity diagrams we draw some conclusions 
on their formation and the history of their evolution. 
\end{abstract}


\section{Introduction}
\label{intro}

The color--magnitude relation (CMR) of early-type galaxies is a well known 
photometric relation that has been studied for long 
\citep*[e.g.][]{B59,VS77,S97,T01,LC04,An06,SC08a,M08,M09,JL09,SC12,L12}. 
In the color--magnitude diagram (CMD), early-type 
galaxies define a sequence in which giant galaxies are redder 
than dwarfs. Spectroscopic and infrared photometric studies of  
early-type galaxies have shown that this relation is mainly driven by 
differences in metal abundances \citep[e.g.][]{Faber73,T99,T00,C11}. 

It is widely accepted that when optical photometric bands are 
considered, the CMR is linear along its extension, and that it shows no 
perceptible change of slope from the bright galaxies to the dwarf regime 
except, sometimes, at the very bright end (e.g. \citealp{S97}, $(B-R)$ vs. 
$R$ in Coma; \citealp{T01}, $(U-V)$ vs. $V$ in Coma; \citealp{LC04}, $(B-R)$ 
vs. $R$ in Coma and 57 X-ray detected clusters; \citealp{An06}, $(B-V)$ vs. 
$R$ and $(V-R)$ vs. $R$ in Abell\,1158; \citealp{SC08a,SC12}, $(C-T_1)$ vs. 
$T_1$ in Antlia; \citealp{M08}, $(V-I)$ vs. $V$ in Hydra; \citealp{M09}, 
$(V-I)$ vs. $V$ in Centaurus). In addition, there is strong 
evidence for the universality of this relation in clusters of galaxies which
led several authors to suggest its use as a reliable indicator of 
distance and cluster membership \citep[e.g.][]{S72,BJ98,LC04,Chib10}.

Despite all the evidence about the existence of a common linear 
CMR in nearby groups and clusters, 
non-linear trends have been reported in Virgo. 
\citet{JL09} and \citet[][hereafter, Ch10]{Chen10} 
have obtained an inverse S-shaped relation that seems to be consistent 
with the quadratic fit performed by \citet[][hereafter, F06]{Ferrarese06}. 
More recently, \citet{L12} have reported a linear
relation that changes its slope at $M_V\approx -14$ mag, where the dSph 
regime begins. 

Based an all these findings, the CMR of Virgo early-type galaxies and,
therefore, Virgo's early-type galaxy population itself, would 
display distinct characteristics in comparison with similar populations
in other groups and clusters. These characteristics should be explained in the
light of theoretical models or/and observational data. 

Bearing this in mind, we revisit Virgo's CMR by performing our own photometry 
and analysis on the images of 100 early-type galaxies included in the 
{\it ACS Virgo Cluster Survey} (ACSVCS) of the Hubble Space 
Telescope (HST). This material has been already used by F06 to analyse this 
relation. Here we present a new analysis based on photometric
parameters obtained with a simpler approach than that followed by F06.  
Basically, we obtained brightness profiles by fitting elliptical
isophotes with fixed parameters, and calculated isophotal limited
magnitudes, colors and effective radii from these brightness
profiles. By comparing our results with those of F06 and those
obtained by Ch10 from data of the {\it Sloan Digital Sky Survey
(SDSS)}, we aim to explore the alleged non-linearity of the relation.  

The paper is organized as follows. Section\,\ref{observations} presents the 
data and our approach to obtain photometric and structural parameters.
Section\,\ref{results} shows our analysis of the color--magnitude and 
mean effective surface brightness ($\langle \mu_{\rm eff} \rangle$)--luminosity
diagrams obtained from the photometry of F06. We also compare these 
diagrams with those built from our photometry and from the photometry 
of Ch10. Section\,\ref{discusion} presents a discussion of the results,
and Section\,\ref{conclusion}, our conclusions.

\section{Observations and photometry}
\label{observations}

From the {\it Multimission Archive at the Space Telescope Science Institute} 
(MAST), we downloaded the images of 100 early-type galaxies confirmed as 
members of the Virgo cluster through radial velocities. These 
images were obtained in the framework of the ACSVCS 
\citep{C04}, and were taken with the F475W and F850LP filters and 
automatically processed with the {\it CALACS} and {\it MultiDrizzle} 
packages \citep{K02}. Short exposures images were obtained in order 
to recover the very central regions of galaxies with saturated nuclei. 
However, we decided to work only with those of long exposures (750 sec. for 
the F475W filter and 1200 sec. for the F850LP one) as the overexposed zones
represent just the inner $0.4''$ in the worst cases. The field of view 
(FOV) of the ACS camera is $202 \times 202'' $ and subtends a spatial 
scale of $0.049''$ pixel$^{-1}$. Following F06, we will adopt a distance 
modulus for Virgo of $(m-M)=31.09$, which translates into a distance of 
$D=16.6$ Mpc. At this distance, $1''$ subtends $\sim$80 pc.

Our photometry, presented in Table\,\ref{Table1}, was obtained 
with ELLIPSE  within IRAF, adopting a simpler approach than that followed by 
F06. As a first step, we constructed a master mask to fill the gaps and cover 
the bad pixels of the ACS CCDs. After that, we built individual masks for each 
galaxy in order to avoid bright or saturated objects and bad or overexposed 
regions. From the galaxies images in the F475W filter, we obtained their 
center coordinates, and from the (F475W-F850LP) color maps, we estimated their 
global ellipticities and position angles as those shown by their external 
regions. Then, we fitted elliptical isophotes keeping these parameters fixed 
during the run. 

Eight galaxies displayed saturated central regions. In these cases, the 
isophotes were fitted excluding these regions, from an initial radius of 
$0.4''$ at most. All of them are bright objects ($M_g\lesssim-18.9$ mag), so 
we do not expect that the excluded regions significantly affect the 
integrated colors and luminosities. 

Once we obtained the brightness profiles in both filters, we examined the 
images one by one in order to identify those galaxies that are not fully 
contained in the FOV of the ACS (hereafter, {\it incomplete galaxies}). 
For those galaxies that are complete in the ACS frames (hereafter, 
{\it complete galaxies}), we corrected the sky level by constructing growth 
curves, i.e., plots of integrated flux versus semi-major axis. The correction 
to the sky level is the value for which these curves display an asymptotic 
flat behavior to infinity. 

After that, we calibrated all brightness profiles to the ABMAG system 
\citep{S05} following F06. Through this calibration, F475W and F850LP surface 
brightnesses transform into $g_{AB}$ and $z_{AB}$ ones, respectively, which 
resembles the SDSS $g$ and $z$ filters.

Integrated SDSS $g$ and $z$ magnitudes, and the corresponding $(g-z)$ colors,
were obtained by integrating the calibrated brightness profiles in each 
filter. For complete galaxies, the integration was performed up to 
$\mu_{g_0}\approx27$ mag arcsec$^{-2}$. For incomplete ones, the 
integration was performed up to a limiting surface brightness of 
$\mu_{g_0}\approx 24$ mag arcsec$^{-2}$, as the corresponding isophotes 
did not reach the edge of the images. We expect that at this brightness the 
profiles are not strongly affected by the adopted sky value: for complete 
galaxies, the sky correction at this level is $\lesssim0.12$ mag 
arcsec$^{-2}$. Regarding the effect of the different surface brightness 
limits on the integrated magnitudes, we find an average difference for 
complete galaxies of $\langle (g_{24}-g_{27})\rangle=1.0\pm0.8$ mag.

Finally, magnitudes and colors, as well as brightness and color profiles, 
were extinction and reddening corrected considering 
$A_g=3.591~E(B-V)$ and $A_z=1.472~E(B-V)$ \citep{S05}. The values of 
$E(B-V)$ for each galaxy were obtained from table\,1 of F06. 

We also estimated effective radii ($r_{\rm eff}$) for the complete galaxies. 
To do this, we considered the total flux of the galaxy as that contained 
within the limiting isophote used to obtain magnitudes and colors, and 
the $r_{\rm eff}$ as the equivalent radius that contains half of that flux. 
From these values, we calculated the corresponding 
$\langle \mu_{\rm eff} \rangle$ following equation 1 of \citet{SC08a}.
{\it In this way, our $r_{\rm eff}$ and $\mu_{\rm eff}$ values are model 
independent}. In addition, it should be noticed that they will probably 
reflect the effects analyzed by \citet{Trujillo01} related with structural 
parameters obtained from images limited in their radial extent 
(for example, it is expected that the $r_{\rm eff}$ of galaxies with 
$n\ge4$ are underestimated by a factor of 2, and their 
$\langle \mu_{\rm eff} \rangle$, overestimated in 1 mag or more).

In the publicly available webpage of our 
group\footnote{http://fcaglp.unlp.edu.ar/CGGE/Virgo/Virgolist.html} we show 
the $F475W$ images, $(g-z)$ color 
maps, surface brightness profiles and color profiles obtained from our 
photometry for the 100 galaxies of the ACSVCS. They are grouped according
to the zone that occupy in the CMD (see Section\,\ref{CMR_F06}), ordered by 
their VCC number. 

\section{Results}
\label{results}

\subsection{Virgo CMR built from the ACS photometry of F06}
\label{CMR_F06}

\begin{figure*}
\center
\includegraphics[scale=0.85]{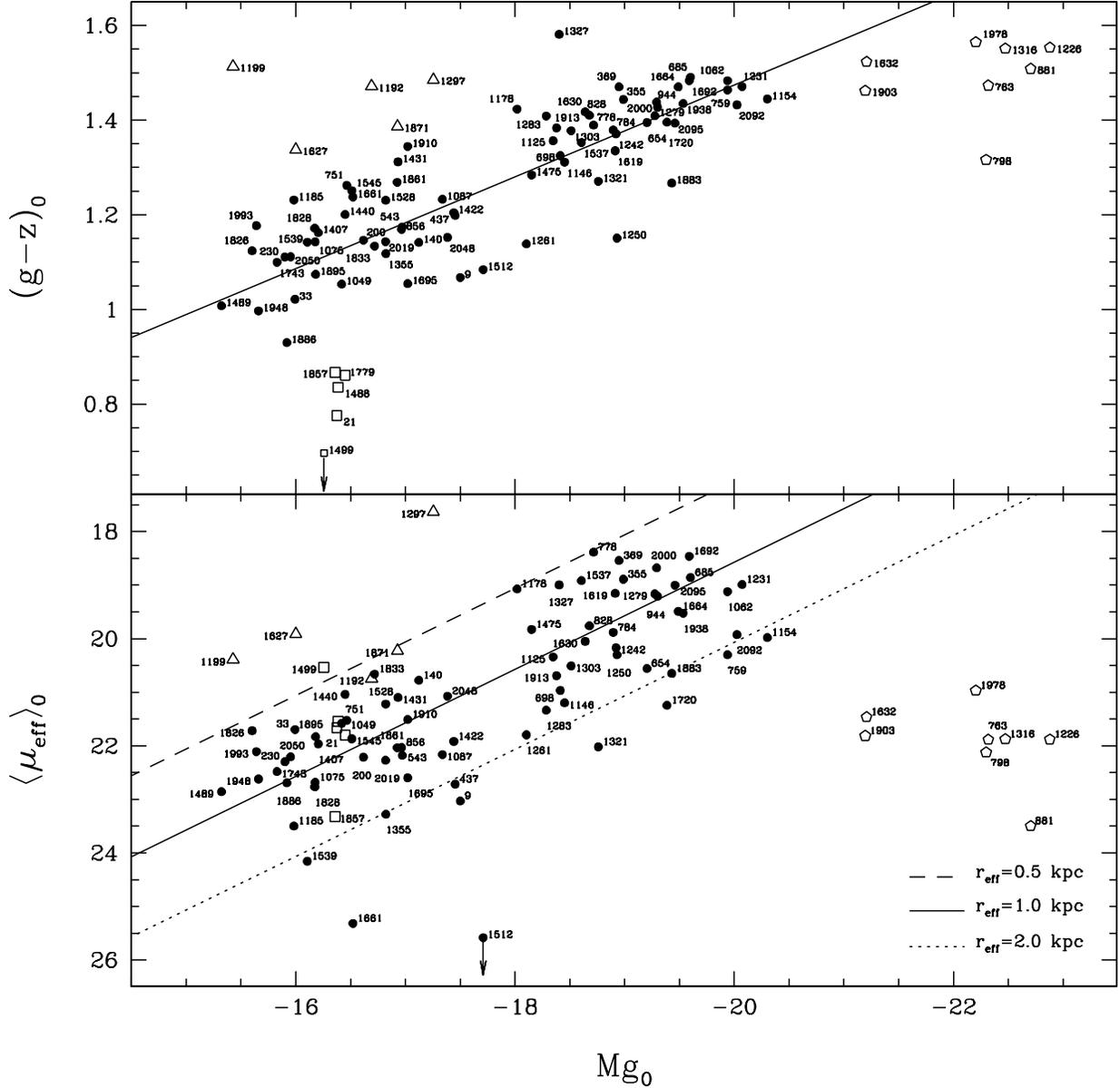}
\caption{Distance, extinction and reddening corrected color--magnitude  
{\it (top)} and $\langle\mu_{\rm eff}\rangle$--magnitude {\it (bottom)} 
diagrams of the galaxies included in the ACSVCS, 
built from magnitudes, colors and $\langle\mu_{\rm eff}\rangle$ values 
reported by F06. The numbers correspond to the VCC designation of each 
galaxy. With open pentagons, we show the galaxies that define a break in the 
CMR towards the bright end. Open triangles depict objects consistent with 
being cE galaxies, and open squares, those showing blue centers and evidence 
of recent star formation. For clarity, in the bottom panel, we have not 
labeled VCC\,1488 and VCC\,1779, two galaxies displaying blue centers. The 
dashed line in the top panel represents the linear fit to the 
CMR when bright ellipticals, cE galaxies and dwarfs with evidence of 
recent star formation are excluded from the sample. The different lines shown
in the bottom panel depict loci of constant $r_{\rm eff}$. }
\label{RCMF}
\end{figure*}

In the top panel of Figure\,\ref{RCMF} we show the CMD 
of the galaxies observed within the ACSVCS. To build this plot, we
used the extinction and reddening corrected $g$ magnitudes and $(g-z)$ 
colors presented by F06 in their table\,4. We recall here that 
these magnitudes exclude the nucleus and are obtained by 
integrating the best-fit profile to infinity, while colors are measured 
directly from the surface brightness profile in the range 
$1'' < r < r_{\rm eff}$. Absolute magnitudes are calculated from 
the distance moduli of the individual galaxies given by \citet{Mei07} 
when available. We exclude from the sample the galaxies 
VCC\,538, VCC\,571, VCC\,575, VCC\,731 and 
VCC\,1025 as \citet{Mei07} found they lie behind the Virgo cluster. 

In the CMD we can distinguish three regions:

\begin{enumerate}
\item A break at the bright end defined by VCC\,763, VCC\,798,
VCC\,881, VCC\,1226, VCC\,1316, VCC\,1632, VCC\,1903 and VCC\,1978. There is 
a gap between this break and the rest of the galaxies, at 
$M_{g_0}\sim -20.6$ mag.
\item A red group depicted by VCC\,1192, VCC\,1199, VCC\,1297, VCC\,1627 and
VCC\,1871.
\item A blue zone defined by VCC\,21, VCC\,1488, VCC\,1499, VCC\,1779 and 
VCC\,1857. 
\end{enumerate}

With the exception of VCC\,1327, the rest of the galaxies define 
what seems to be a linear trend with a gap at $M_{g_0} \sim -17.5$ mag 
(hereafter, {\it intermediate gap}). This gap is probably
due to an incompleteness effect as the ACSVCS sample is $\sim 60\%$  
complete, while it is $100\%$ complete for galaxies brighter
than $M_{g_0} \sim -19.5$ mag (Ch10). 

Taking into account the errors given by F06 for $(g-z)$ colors, we  
performed several fits to the linear trend. In Table\,\ref{ajustes} we show 
the values obtained for the different parameters. There seems to be a change 
of slope between the bright ($M_{g_0}\lesssim -17.5$) and faint 
($M_{g_0}> -17.5$) parts of the linear relation. However, this is 
not a reliable result as the sub-samples considered to perform these fits 
display weak linear correlations ($r<-0.6$). This weakness is also detected 
in the analysis of the intrinsic scatter of the relation. When the number of 
galaxies considered to perform the bright and faint fits vary, we obtain 
contradictory results for $\sigma_{(g-z)_0}$ (it increases towards the faint 
end in the whole sample, while it decreases in the reduced one).

However, it can be stated that the dispersion in the relation might not
be due to distance effects. The mean measurement error in the distance 
moduli reported by \citet{Mei07} is 0.07 mag. We have found that the 
intrinsic scatter in magnitude around the mean relation is 0.82 mag.  

When we build the $\langle \mu_{\rm eff} \rangle$--luminosity diagram from 
the data obtained by F06 (bottom panel of Figure\,\ref{RCMF}), we can 
distinguish three different zones that deviate from a linear 
color--magnitude trend. The very bright galaxies defining the bright break 
in the CMD also define a break from the general trend followed by most of the 
galaxies. The objects of the red group set a low boundary in $r_{\rm eff}$ 
and a high one in $\langle \mu_{\rm eff} \rangle$ to the whole sample. 
The galaxies that depict the linear trend in the CMD are found around
the locus of constant $r_{\rm eff}\sim1$ kpc, with most of them 
displaying $0.5 < r_{\rm eff} < 2.0$ kpc. However, the galaxies 
that show the bluest colors are placed within the trend of 
dwarf early-type galaxies.   

From the images, color maps and brightness profiles of the galaxies,
it can be seen that the objects defining the bright break in the 
CMD are very bright ellipticals and S0s. In particular, 
this sub-sample includes the three dominant Virgo galaxies VCC\,881 (M86), 
VCC\,1226 (M49) and VCC\,1316 (M87). 

Within the red group, all of the galaxies display 
characteristics typical of compact elliptical (cE) galaxies. That is, 
round and compact morphologies, red colors for their luminosities and 
concentrated brightness profiles ($r_{\rm \mu_{z_0}=25} \leq 40''$) with 
high central surface brightness ($\mu_{z_0}\approx 16$ mag arcsec$^{-2}$). In 
particular, VCC\,1297 is the well known cE galaxy of the Virgo cluster, 
NGC\,4486B. Our identification is in agreement with that of Ch10 
regarding VCC\,1192, VCC\,1199, VCC\,1297 and VCC\,1627. We add the galaxy 
VCC\,1871.

Most of the galaxies defining the blue zone display blue centers that can be 
observed both in the color maps and in the color profiles. The exception is 
VCC\,1857 which does not show a significant color gradient but it is quite 
blue ($(g-z)_0\sim 0.9$ mag). \citet{Mei07} were not able to estimate distance 
moduli for most of these galaxies through the surface brightness 
fluctuation method because of their extremely blue colors. 
 
Among the galaxies that define the linear trend, we noticed
that VCC\,1512 is clearly inconsistent with the size and 
morphology displayed by the rest of the systems of similar integrated 
magnitudes. F06 have reported $r_{\rm eff}=497.5''$ but
they pointed out that all of its integrated quantities should be considered 
unreliable. In the case of the galaxies that are behind Virgo, a similar 
situation is found for VCC\,575 as F06 have reported 
$r_{\rm eff}=490''$.

\subsection{Virgo CMR built from our ACS photometry}
\label{ACS}

\begin{figure*}
\center
\includegraphics[scale=0.85]{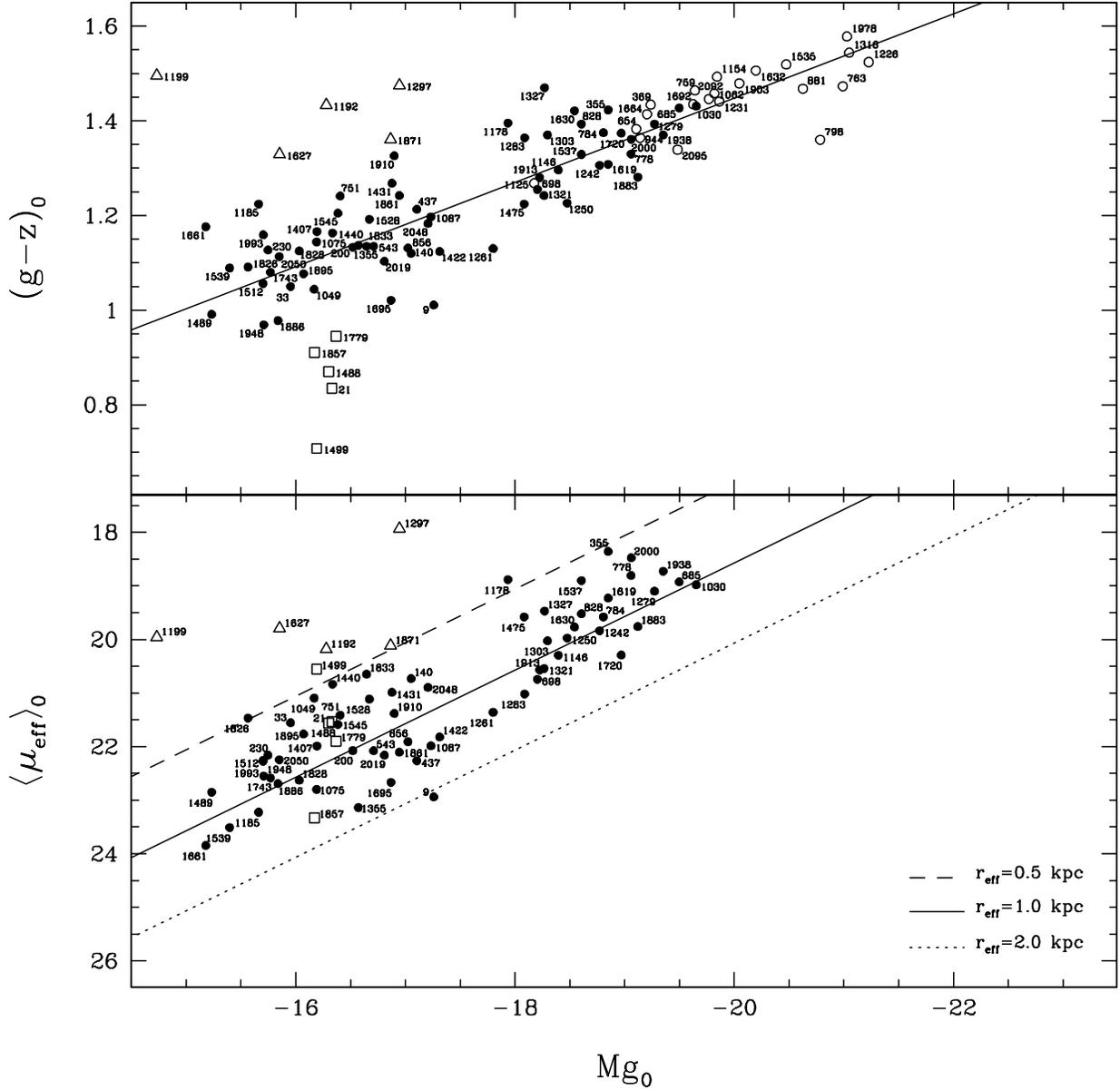}
\caption{Distance, extinction and reddening corrected color--magnitude 
{\it (top)} and $\langle\mu_{\rm eff}\rangle$--magnitude {\it (bottom)} 
diagrams of the galaxies included in the ACSVCS, built from our photometry. 
The numbers correspond to the VCC designation of each galaxy. In the top 
panel, we depict with open circles incomplete galaxies, not shown in 
the bottom panel as we were not able to estimate their $r_{\rm eff}$. With 
open triangles we show the objects consistent with being cE galaxies, and 
with open squares those that show blue centers and evidence of recent star 
formation. The dashed line in the top panel represents the linear fit to the 
CMR when incomplete galaxies, cEs and dwarfs with evidence of recent star 
formation are excluded from the sample. The different lines shown in the 
bottom panel depict loci of constant $r_{\rm eff}$. }
\label{RCMN}
\end{figure*}

In the top panel of Figure\,\ref{RCMN} we show the CMD of 
the galaxies included in the ACSVCS obtained from our photometry. It is 
worth noticing that VCC\,1327 follow the general trend 
in this case. As noticed by F06, there is a bright foreground 
star near the center of the galaxy. To obtain VCC\,1327 brightness and 
color profiles, we masked the star. In addition, we have obtained 
photometric parameters for VCC\,1030. F06 do not give an estimate 
of magnitudes, colors or $r_{\rm eff}$ values for this galaxy. Therefore, we 
include these two objects in our analysis. For VCC\,1535, as it is 
an incomplete galaxy, we obtained low limits for its integrated 
magnitudes and $(g-z)$ color. F06 have not reported integrated quantities or 
structural parameters for it. 

In spite of these differences and those analyzed in 
Appendix\,\ref{valores}, we reproduce the general trend of the CMR. 
In particular, the linear region with its intermediate gap, as well as 
the red and blue zones. As expected, we lose the bright break as the galaxies 
that define it are incomplete ones and the magnitudes obtained from 
our photometry are underestimated. 

In Table\,\ref{ajustes} we show the results of performing different fits
to the linear trend excluding incomplete, cE and blue galaxies. We can 
see that, as in the case of the F06 CMD, there seems to be a change of 
slope between the bright and faint end of the trend, although in the 
opposite way. However, both slopes are comparable within the errors 
in spite of the 
weak linear correlation of the sub-samples. In addition, we have found a 
slight increase of the intrinsic scatter towards the faint end of the 
relation. 

It should be noticed that, in our case, we considered the errors in both 
variables to perform the different fits, while we could only take into 
account the errors in $(g-z)$ in the case of F06 as these authors do not 
provide magnitude errors.   

We would like to point out that if we consider exactly the same galaxy 
sample both for F06 and our photometry ({\it Reduced 
Sample} in Table\,\ref{ajustes}), we obtain, within the errors, the same 
slope and zero point for the relation. These values are also
consistent with those obtained for the whole trend in each case. In addition,
the apparent change of slope due to the low linear correlation of the
samples is also recovered. 

Finally, we consider our sample with magnitudes and colors integrated 
between $1''$ and 1 $r_{\rm eff}$, excluding incomplete, cE and blue 
galaxies. In this case we obtain:
\begin{equation}
(g-z)_0=1.37(\pm0.014)-0.096(\pm0.007)(M_{g_0}+18)
\end{equation}
with $r=-0.87$ and $\sigma_{(g-z)_0}=0.07$. It is interesting to
note that the slope of the relation does not significantly change by 
considering a different radius range to perform color and magnitude 
integrations, and it is identical within the errors to those found in the
photometry of F06. However, the zero point turns out to be redder which is 
consistent with the fact that, in this case, our magnitudes and colors are 
measured within 1 $r_{\rm eff}$.

It is remarkable that as a consequence of our measurement procedure, 
the central regions of the brightest incomplete galaxies show 
magnitudes and colors that suit so well in the linear trend. We emphasize 
that these galaxies were not included in the sample considered to perform 
the fits as their integrated magnitudes were obtained by integrating their 
brightness profiles up to $\mu_{g_0}\approx24$ mag arcsec$^{-2}$. As can 
be seen from the color profiles of the galaxies of the bright end, this 
brightness is reached below 1 $r_{\rm eff}$. Therefore, we find that in the 
CMD, the central regions of bright early-type galaxies seem to follow the 
linear trend defined by fainter ones.

We also built the $\langle\mu_{\rm eff}\rangle$--luminosity plane with 
the complete galaxies. We did not include the incomplete ones because we 
could not estimate their $r_{\rm eff}$. Despite the differences between 
our $r_{\rm eff}$ values and those reported by F06 for the complete galaxies 
(see Appendix\,\ref{valores}), we recover similar behaviors. That is, 
the galaxies of the red cloud set upper and lower limits in 
$\langle\mu_{\rm eff}\rangle$ and $r_{\rm eff}$, respectively; the 
galaxies that define the linear trend tend to be located around the locus of 
constant $r_{\rm eff}\sim1$ kpc; and the bluest galaxies are placed on the 
trend of typical early-type dwarf galaxies.

\subsection{Virgo CMRs built from the SDSS photometry of Ch10}
\label{Chen_RCM}

To support our results we worked with the SDSS 
$g_{\rm COG}$ vs. $(g-z)_{\rm t}$ photometry of Ch10 and the 
distance moduli of \citet{Mei07}. 
We can see from Table\,\ref{ajustes} that the same reduced sample  
considered above, gives identical parameters to those of F06 and our 
work, within the errors. Furthermore, if we take the whole 
ACSVCS sample excluding the bright break, cE and blue galaxies, but with 
the photometry of Ch10 (79 objects), we obtain similar results to
those found from our photometry. 

\begin{figure*}
\center
\includegraphics[scale=0.40]{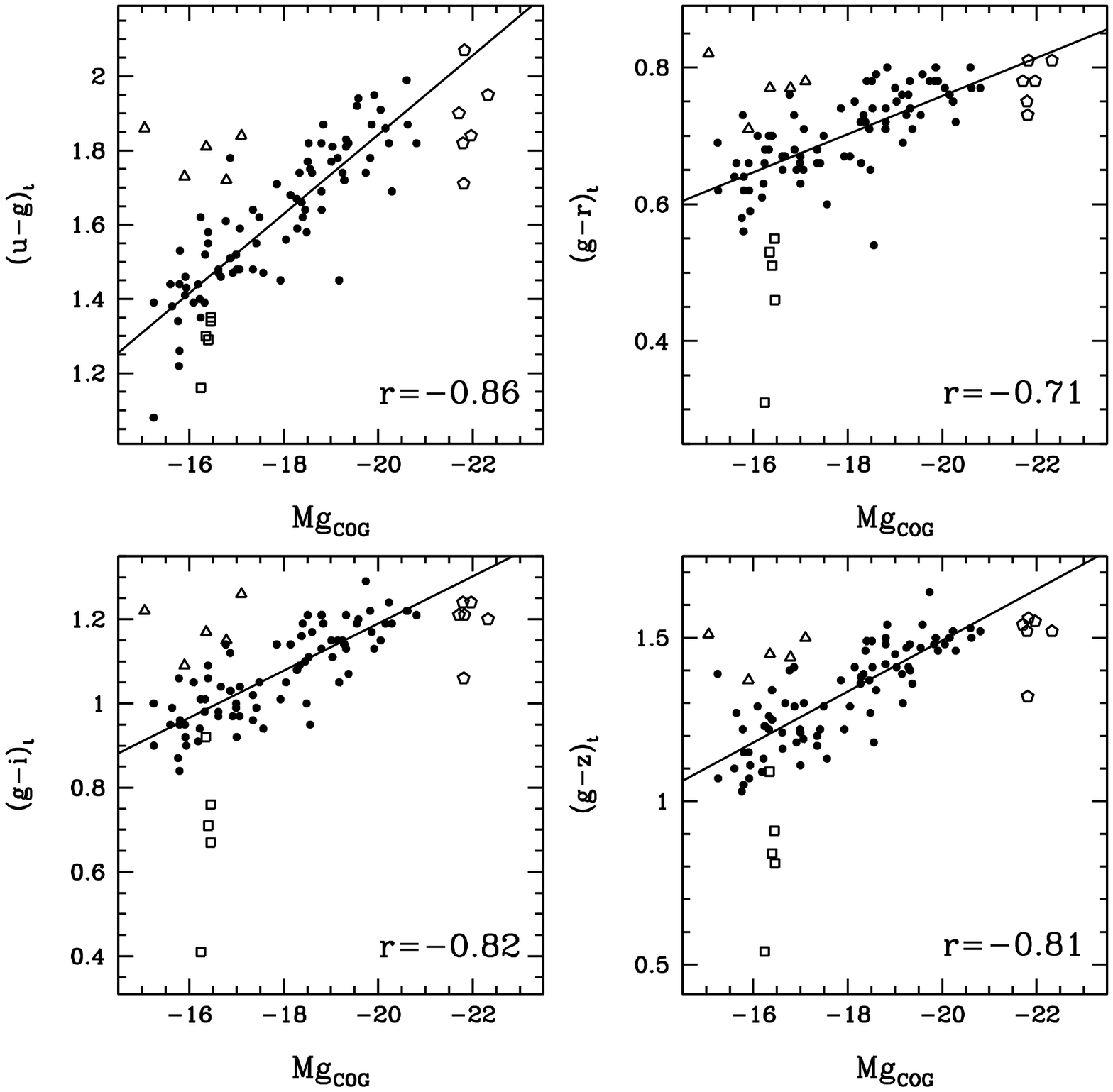}
\includegraphics[scale=0.40]{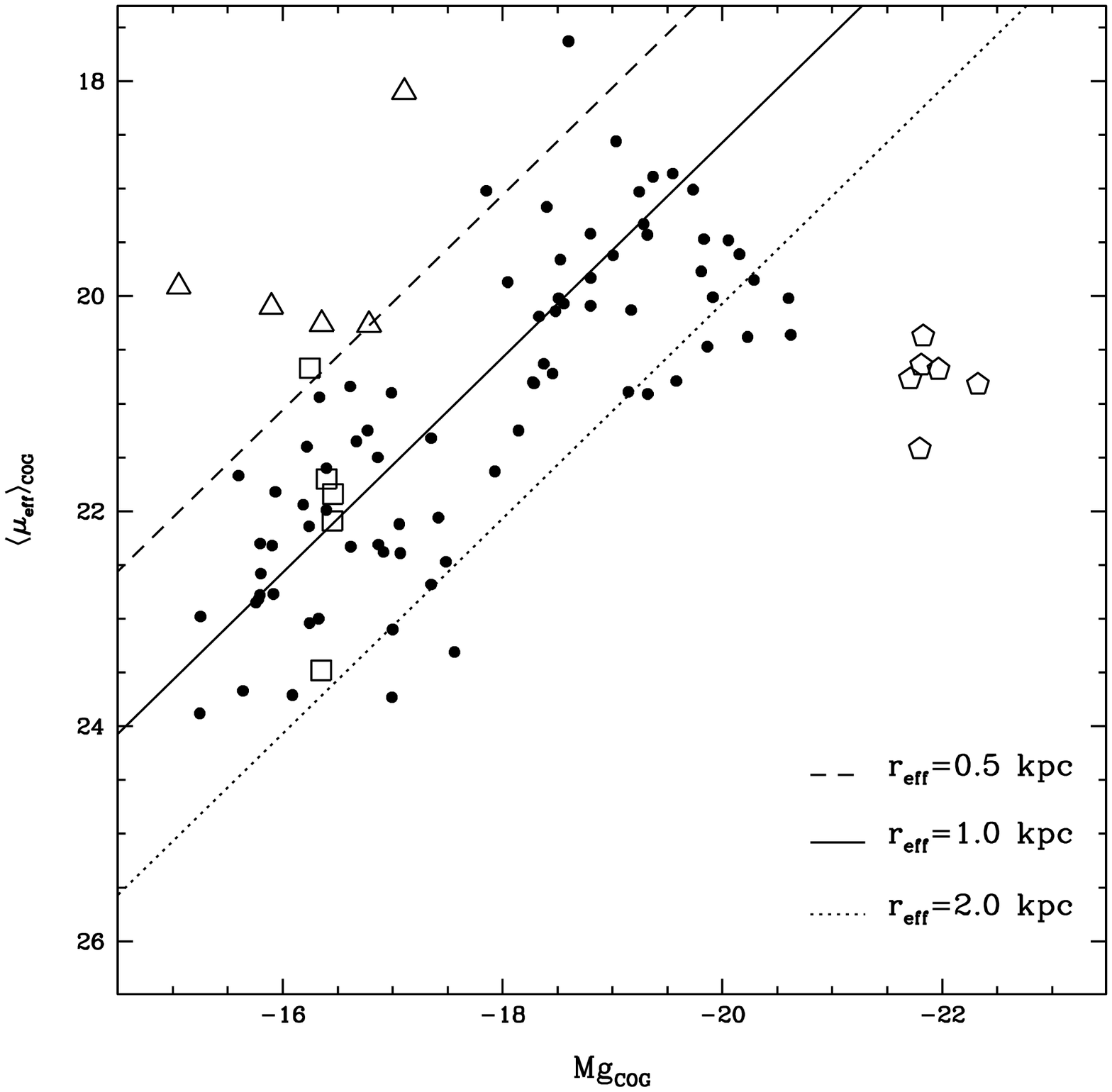}
\caption{{\it Left:} Extinction and reddening uncorrected CMRs obtained 
from the SDSS photometry of Ch10. Symbols code is identical to 
that of Figure\,\ref{RCMF}. The solid lines represent the linear fits 
performed to early-type galaxies, excluding those of the bright break, cE 
galaxies and objects with evidence of 
recent star formation. The Pearson correlation coefficient $r$ found for 
each fit is shown in each panel. {\it Right:} 
$\langle\mu_{\rm eff}\rangle$--magnitude diagram built from the data of 
Ch10. The different lines depict loci of constant $r_{\rm eff}$.} 
\label{Chen}
\end{figure*}

We also obtain strong linear relations when other SDSS color combinations are 
considered, as can be seen from the left panel of Figure\,\ref{Chen}. The 
right panel shows the $\langle\mu_{\rm eff}\rangle$--luminosity diagram built 
with the values presented in table\,1 of Ch10. It is similar to that of F06,
and is similar to our diagram if we exclude the incomplete galaxies.

\section{Discussion}
\label{discusion}
\subsection{The bright break of the CMR}

The bright break in the CMR has been observed in, for example, the Hydra 
cluster. Although \citet{M08} have fit a linear relation to the whole sample, 
a detachment from the general trend in the very bright end is evident from
their figures 8b and 10. It can also be marginally detected in the Centaurus
cluster, where the break is defined by the two brightest galaxies \citep{M09} 
and even in the Coma CMR obtained by \citet{S97}. The break is also seen in 
the CMRs of the Virgo cluster obtained by \citet{JL09} and Ch10 with data 
from the SDSS. It should be noticed, however, that in the CMR of
Ch10, the bright break is only defined by six galaxies (VCC\,763,
VCC\,798, VCC\,881, VCC\,1226, VCC\,1316 and VCC\,1978), while
that of F06 is defined by eight objects. This discrepancy is probably 
due to the difficulty of measuring large galaxies in the small 
FOV of the ACS. In this sense, it is worth noticing the clear 
differences between the values of the $r_{\rm eff}$ found by F06 and 
Ch10 for the galaxies of the bright break.  

From the point of view of numerical simulations, \citet{J11} have shown that 
the bright break in the CMR may arise due to the contribution of dry mergers 
to the final mass of the brightest galaxies of clusters. Dry mergers involve 
little amount of gas and, therefore, they do not trigger starbursts that can 
modify the metal content of the galaxies. On the contrary, they only 
contribute to increase the stellar mass and, therefore, the total luminosity 
at constant metallicity (i.e. color). The bright break in these simulations is 
found at $M_R\sim M_V \sim M_{T_1} \approx -20$ mag. In the CMR of F06, it is 
observed at $M_{g_0} \sim -20.6$ mag and in that of Ch10, at 
$M_{g_0} \sim -21.5$ mag.

On the other hand, \citet[][and references therein]{Hilz12} have found that 
stripped material from minor mergers is expected to alter the external 
regions of the galaxies rather than the inner ones, increasing the size of 
compact massive spheroidal galaxies similar to those observed at $z\sim2$. In 
this context, our findings about the central regions of massive galaxies 
behaving as low-mass early-type galaxies in the CMD, could be 
linked with this picture. If the accreted low-mass satellites were metal-poor 
(thus blue), they would make the global color of the growing galaxies bluer 
than the linear CMR \citep[see also][]{Hilker99,Cote2000}.  

It is interesting that the galaxies defining the bright break in the CMR, also 
detach from the general trend followed by the rest of the galaxies in the 
three-dimensional plot presented by \citet*{F09}. This plot represents a 
logarithmic volume space defined by stellar mass, projected stellar mass 
density and total GC efficiency of formation. The authors conclude that the 
different behavior among the brightest galaxies and the rest of the sample 
may be associated with different merger histories. For these authors, galaxies 
with masses below $10^{11}~M_\odot$ might have experienced a relatively low 
number of dissipationless merging events while systems with masses above that 
value, might mainly have experienced major mergers.

\subsection{The linear region of the CMR}

From our analysis we have found that complete non-peculiar early-type 
galaxies with luminosities in the range $-20.5\lesssim M_{g_0}\lesssim -16.5$ 
mag, which are confirmed members of the Virgo cluster, define a linear 
trend in the CMD. It is worth noticing that from the ACSVCS sample, it 
can only be stated that the CMR of Virgo is linear in the above range which 
covers four magnitudes in brightness. As a comparison, the CMR of the Antlia 
cluster covers $\sim 10$ magnitudes considering spectroscopically confirmed 
members \citep{SC12}. 

The linear region is also recovered in the CMD built from the SDSS photometry 
of the ACSVCS galaxies of Ch10, as it was shown in Section\,\ref{Chen_RCM}. 
They report an inverted S-shape trend in agreement with \citet{JL09}. However, 
this feature arises as an attempt to reproduce the whole trend, including the 
brightest galaxies, the cEs and those of the blue zone. In particular, the 
latter galaxies have been considered in previous works as dE/dIrr transition 
objects which, in our opinion, should be excluded from an early-type galaxy 
sample. 

The analysis of F06 of the relation is quite similar to our work as 
they do not include most of the galaxies of the red group and the blue zone. 
However, to fit the whole relation including the very bright end, 
requires, at least, a second order solution. 

Regarding the galaxies that populate the faint end of the CMD, which
mostly define the inverted S-shape reported by \citet{JL09}, a point to take 
into account is, for example, the different image resolution between 
the data obtained from HST-ACS and SDSS (a factor of $\sim 15$, 
Ch10). Bearing this in mind, it cannot be ruled out that the 
sample used by \citet{JL09} includes a considerable amount of background 
galaxies that would populate the whole CMD (see figure 2 
in \citealt{SC12} for a visualization of this effect), and dE/dIrr objects 
placed in the blue side of the early-type relation. All these systems would
contribute to depict an S-shaped feature.

In the case of the CMR presented by \citet{L12}, the ACSVCS does not reach 
galaxies as faint as that study ($M_{\rm V}\sim-8.5$ mag; using 
table\,3\,{\it(m)} in \citet{F95}, this translates into $M_{\rm g}\sim-8.1$ 
mag). It might be expected that at the very low-mass end, the relation 
vanishes due to the physical limit of the metallicity of a galaxy. Here,
age differences also become a relevant effect in producing the observed 
scatter. In addition, in these faint luminosities we are faced with large 
photometric and member identification errors. However, it should be noticed 
that in their figure\,5, all but three of the spectroscopically confirmed 
Virgo members considered in the analysis, seem to define a quite linear trend 
spanning $-20\lesssim M_{\rm V}\lesssim-16$ mag, which transforms into 
$-19.6\lesssim M_{\rm g}\lesssim-15.6$ mag. This range is quite similar to 
that covered by the linear trend of the ACSVCS sample.

It is interesting to compare the slopes of Virgo's CMR with those found for 
other clusters in the literature. To this aim, and under the hypothesis that 
early-type galaxies colors are mainly dominated by old stellar populations, we 
can use several color--color relations presented in different papers. 

In Virgo we have found $\Delta (g-z)/\Delta g\sim-0.09$. From the color--color 
relations presented by \citet{F13} in their table\,4, this translates into 
$\Delta (C-T_1)/\Delta T_1\sim-0.11$.
In the Antlia cluster, \citet{SC12} find $\Delta T_1/\Delta (C-T_1)=-14.2$ 
when only elliptical confirmed members that do not display cE morphologies 
are considered. The inverse slope gives $\Delta (C-T_1)/\Delta T_1\sim-0.07$.

In the Hydra cluster, \citet{M09} find $\Delta (V-I)/\Delta V=-0.040$. Through
eq. (2) and (3) of \citet{SC08a}, this
transforms into $\Delta (C-T_1)/\Delta T_1\sim-0.10$.  A similar result is
obtained for Centaurus as it displays 
$\Delta (V-I)/\Delta V=-0.042\pm0.001$ \citep{M09}. For Fornax, a shallower
slope is found ($\Delta (V-I)/\Delta V=-0.033\pm0.004$, \citealp{M08}) 
which translates into $\Delta (C-T_1)/\Delta T_1\sim-0.08$. \citet{S97} find 
for Coma $\Delta (B-R)/\Delta R=-0.056$ which becomes 
$\Delta (C-T_1)/\Delta T_1\sim-0.08$. 

From the above analysis we obtain that Fornax, Antlia and Coma seem to present
a similar slope, shallower than those of Virgo, Hydra and Centaurus, 
which agree among them. In this sense, it is interesting to note 
that \citet{L12} have found a linear CMR for Virgo with a 
slope $\Delta (V-I)/\Delta V=-0.045\pm0.007$. This value, compared
with those obtained in the same photometric system for Hydra, Centaurus and 
Fornax, is, within the errors, similar to those of Hydra and 
Centaurus, but steeper than that of Fornax, in agreement with our 
findings. In addition, Virgo and Centaurus seem to show similar
values for the zero point of the relation.

\subsection{cE galaxies}

The location of cE galaxies in the color--magnitude and 
$\langle\mu_{\rm eff}\rangle$--luminosity diagrams of Virgo are in 
agreement with those of cE galaxies in the Antlia cluster. That is, they are 
the galaxies that display the reddest colors, smallest $r_{\rm eff}$ and 
brightest $\langle\mu_{\rm eff}\rangle$ for their luminosities in a sample of 
early-type galaxies. The behavior of cEs defining a trend displaced from that 
followed by dEs in the $\langle\mu_{\rm eff}\rangle$--luminosity plot rather 
than extending the region occupied by luminous galaxies, is also consistent
with that shown by \citet{Chilingarian09} for cE galaxies found through the
Virtual Observatory. In consequence, the color--magnitude and 
$\langle\mu_{\rm eff}\rangle$--luminosity diagrams complement each other
giving a photometric tool to identify cE candidates.

As there are more galaxies of this kind in Virgo than, for example, in 
Antlia, we wonder if it is possible to link the position of cE galaxies in 
the color--magnitude and $\langle\mu_{\rm eff}\rangle$--luminosity diagrams 
with some of their features in order to give clues about their 
physical origin. 

For example, those displaying the reddest colors are VCC\,1192, 
VCC\,1199 and VCC\,1297. Among them, VCC\,1199 and VCC\,1297 show the smallest 
$r_{\rm eff}$ of the cEs sample, while VCC\,1192 shows a size more similar 
to those of dE galaxies. This might be taken as evidence that 
VCC\,1199 and VCC\,1297 have experienced a stronger interaction with their 
massive neighbors than VCC\,1192, generating their  
compactness but not necessarily their present integrated colors, as the three 
galaxies are comparable in this sense. However, VCC\,1192 and VCC\,1199
are found, in projection, closer ($\sim$4 arcmin) to M\,49, than VCC\,1297 
to M\,87 ($\sim$ 7.5 arcmin). And the relative radial velocities of
the formers to M\,49 ($\sim$400 km s$^{-1}$) are larger than that of the 
latter to M\,87 ($\sim$200 km s$^{-1}$). 

In contrast, VCC\,1627 and VCC\,1871 display the bluest colors of the 
cEs sample. The $r_{\rm eff}$ of VCC\,1627 is smaller than that of VCC\,1192,
while that of VCC\,1871 is larger. VCC\,1627 is placed, in projection, at
$\sim$10.5 arcmin from M\,89 with a relative radial velocity of $\sim$ 100
km s$^{-1}$, and at $\sim$7.5 arcmin from NGC\,4550 with a relative velocity 
of $\sim$140 km s$^{-1}$. VCC\,1871 is found at $\sim$19 arcmin from M\,59 
with a relative radial velocity of $\sim$180 km s$^{-1}$. 

Therefore, at the moment, it is difficult to realize if there is a 
relationship between the spatial locations of cE galaxies and their colors or 
mass distribution, and to try to explain their unusual features. There is 
growing evidence of these objects being the remnants of disrupted systems as 
a consequence of their interaction with massive galaxies 
\citep[see, for example,][]{SC08b,Chilingarian09,Huxor11}. Some of our 
findings could be pointing towards a link between the present colors of the 
objects and their progenitors, and between their degree of compactness and the 
eccentricity of the orbits around their brighter companions. However, a more detailed 
study is necessary before arriving at firm conclusions. 

\subsection{The color--magnitude and $\langle\mu_{\rm eff}\rangle$--luminosity 
diagrams link}

There seems to be a clear connection between the zones identified 
in the CMD of Virgo with those observed in the 
$\langle\mu_{\rm eff}\rangle$--luminosity one. While faint and 
intermediate early-type galaxies define linear trends in both diagrams, 
the brightest galaxies and cEs depart from them in both plots. The exception 
are the galaxies showing evidence of recent star formation that follow the 
locus of early-type dwarfs in the $\langle\mu_{\rm eff}\rangle$--luminosity 
diagram but clearly depart from the linear region of the CMR. This fact 
shows that if low-mass galaxies showing recent star formation are included 
in an early-type galaxy sample only because of their morphology or structural 
parameters, can cause significant contamination. 

If we consider that the CMR represents a mass--metallicity relation, and the 
$\langle\mu_{\rm eff}\rangle$--luminosity diagram provides structural 
information, from these plots we can detect the link between 
structural and chemical properties in early-type galaxies. For example,
all the galaxies that define the linear region of the CMR, 
display $r_{\rm eff}$ in a restricted range of $\sim 0.5 - 2.0$ kpc 
\citep{SC08a,M11}, which results in an increase of their 
$\langle\mu_{\rm eff}\rangle$ with luminosity. On the contrary, the brightest 
and cE galaxies define the highest and lowest boundarys in $r_{\rm eff}$, 
respectively, displaying no clear trends in the CMD: cE galaxies show 
$r_{\rm eff} < 0.5$ kpc, and the brightest ones, display 
$r_{\rm eff} \gtrsim 8$ kpc. 

It is interesting to note that the connection between the different
zones in the CMD and $\langle\mu_{\rm eff}\rangle$--luminosity plane 
are also seen in other galaxy clusters, but with a different level of 
definition. The linear regions of both diagrams are easily detected 
in several clusters. However, the bright break and the population of cE 
galaxies are revealed depending on the group or cluster. 
For example, in Antlia there seems to be a break in the 
bright regime of the $\langle\mu_{\rm eff}\rangle$--luminosity diagram, but 
there is no apreciable break in the CMR. In addition, only two cE 
galaxies have been identified until today. In the Hydra cluster, only two of the four 
galaxies that seem to define the bright break in the CMR, define a clear 
departure from the general trend in the 
$\langle\mu_{\rm eff}\rangle$--luminosity plot, and there are no cE galaxies 
reported in this cluster.  

Excluding dE/dIrr transition objects, both diagrams seem to point towards 
different evolutionary paths for early-type galaxies. 
The brightest galaxies may mainly assemble via mergers and/or accretion of 
low-mass satellites while cEs might arise from the disruption of more massive 
systems. On the other hand, intermediate and low-mass early-type galaxies 
would form {\it in situ} and passively evolve in a way that encounters and 
mergers do not significantly alter their global evolution. 
The fact that some but not all CMRs show a bright break would be in agreement 
with the merger/accretion processes taking place at different 
times and/or levels in different clusters. In a similar way, disruption 
events will generate objects with different degrees of compactness 
depending on the progenitor, the strength of the interaction and/or the stage 
in time of the disruption process. A mainly passive evolution would 
produce well defined and ordered relations.

This picture, and that of central regions of galaxies following the 
linear trend of the CMR, fits quite well in the galaxy formation scenarios 
analyzed by \citet{Oser10} with numerical simulations. In these models, 
high-mass galaxies ($M_{\star} \lesssim 1.7 \times 10^{11}~ M_{\odot}~ 
h^{-1}$) assemble mainly by accretion and merging, with about 80\% of the 
present--day stars added by these events and building a more extended 
component, while low mass galaxies 
($M_{\star} \lesssim 9 \times 10^{10}~ M_\odot~h^{-1}$) form up to 60\% of 
their stellar mass {\it in situ}. These authors find that at high redshift, 
the assembly of galaxies of all masses is dominated by {\it in situ} star 
formation fed by cold flows. After this, the importance of stellar accretion 
increases with galaxy mass and, for the most massive galaxies, the 
accretion process is an important event at low redshifts.    

The very faint end of both diagrams represents a more difficult task for 
interpretation because photometic errors, morphological 
missclassification, background contamination or, even, isophotal truncation 
(see figure\,3 in \citealp{SC12}) can blur intrinsic trends. 
Before proposing an hypothesis for their origin in an observational 
basis, an effort is needed to spectroscopically confirm more low-mass cluster 
members. In this sense, it is worth noticing that, in the literature, CMDs 
of galaxy clusters are built with confirmed members up to $M_{\rm V}\sim-13$ 
mag as most \citep{M07,M08,M09}. The alleged faint break in the Virgo cluster 
is not observed at this level for confirmed members. Only faint galaxy 
candidates seem to define such a break. In addition, \citet{D09} have built 
the CMD of dE/dSph of different groups and clusters in a range of 8 mag, 
up to $M_{\rm V}\sim-10$ mag, and all follow the same tight linear 
relation extrapolated from that defined by the brightest galaxies.

\section{Conclusions}
\label{conclusion}

In order to deepen our understanding of the CMR of early-type galaxies,
we revisit the CMR of the Virgo cluster by analyzing that obtained
by F06, and performing our own photometry of the 100 galaxies included in 
the ACSVCS. Although we obtain integrated colors and magnitudes with a 
simpler approach, we recover the general trend of the CMR reported by F06, 
except at the very bright end due to the surface brightness limit 
adopted for the spatially incomplete galaxies.

We have found that when galaxies displaying the highest luminosities,  
compact morphologies or evidence of recent star formation are excluded from the 
sample, a linear trend is recovered with a gap at $M_{g_0}\sim -17.5$ mag, 
which is probably due to incompleteness effects. Therefore, it is not necessary 
to fit quadratic or higher order functions to the CMR of the Virgo cluster in 
the magnitude range $-20.5\lesssim M_{g_0}\lesssim -16.5$ mag.  Furthermore,
the linear trend in this luminosity range is strong enough to be recovered
with different color combinations and simple photometric methods, even  
considering a lesser number of galaxies .

Other Virgo CMRs reported in the literature are consistent with our analysis 
once a detailed morphological classification and/or membership identification 
is taken into account. Also, the slope of the CMR of Virgo has been compared 
with those of other galaxy clusters. Within the errors and considering 
the difficulty of comparing quantities obtained in different photometric 
systems, our results are also consistent with those of Hydra, and 
Centaurus, and it is steeper than those of Fornax, Antlia and Coma, which 
present similar slopes.

It is remarkable that the central regions of the brightest galaxies follow the 
linear trend depicted by the rest of the galaxies, rather than defining a 
stochastic distribution in the CMD due to their underestimated brightnesses. 
This finding might suport the picture in which large and massive elliptical 
galaxies observed today, are formed from compact and massive spheroids that enlarge 
their sizes through mergers and/or accretion of low-mass satellites.

We have identified four cE galaxies besides the well known VCC\,1297 
(NGC\,4486B). Some of these galaxies have been previously reported as very 
compact and round systems, but we emphasize their similarity with cEs 
detected in other groups and clusters. They occupy very specific 
regions both in the color--magnitude and 
$\langle\mu_{\rm eff}\rangle$--luminosity diagrams: they are the reddest and 
most compact objects for their luminosities in a sample of early-type 
galaxies. In this sense, both diagrams analyzed together might be used to 
detect cE galaxy candidates.

A clear connection can be made between the different regions identified in the 
color--magnitude and $\langle\mu_{\rm eff}\rangle$--luminosity diagrams of 
Virgo. The galaxies that define the bright break in the CMR, also 
detach from the rest of the systems in the 
$\langle\mu_{\rm eff}\rangle$--luminosity diagram; the galaxies that follow 
the linear trend of the CMR also define a locus around a constant 
$r_{\rm eff}\sim1$ kpc, displaying $r_{\rm eff}$ between 0.5 and 
2.0 kpc; and the cE galaxies which define a color limit for their 
luminosities in the CMD, also define $\langle\mu_{\rm eff}\rangle$ and 
$r_{\rm eff}$ limits in the $\langle\mu_{\rm eff}\rangle$--luminosity plot. 

The connection between the different zones occupied by different
early-type galaxy sub-samples in the color--magnitude and 
$\langle\mu_{\rm eff}\rangle$--luminosity diagrams could be taken as the 
manifestation of the link between color (abundance) and structural 
properties. In addition, it might be considered as evidence of the relative 
importance played by different formation mechanisms in different 
sub-samples of early-type galaxies. Object disruption seems to be an 
important process behind cEs' evolution while mergers and/or satellite 
accretion might be the main drivers in the case of the very bright early-type 
galaxies. As the final products of such phenomena depend on, for example, the 
type and/or relative masses of the progenitors, it would be expected that they 
translate into no defined trends in both diagrams. On the contrary, a more 
passive evolution in which interactions play a less important role, might 
dominate the evolutionary path of intermediate and low-mass early-type 
galaxies producing well defined relations along an important range of 
masses.

If this scenario was correct, passive evolution would be a quite 
common process as the linear trends in the CMD 
and $\langle\mu_{\rm eff}\rangle$--luminosity planes are observed in 
different groups and clusters. In this context, if the differences found in 
the slopes and zero points of some CMRs were significant, they could be 
interpreted as the manifestation of this process taking place with, for 
example, different enrichment rates or/and efficiencies in distinct 
associations. 

The exception are the blue galaxies, which depart from the linear
region of the CMR but follow the locus of normal early-type galaxies in the 
$\langle\mu_{\rm eff}\rangle$--luminosity diagram. In this sense it should be 
noticed that their asymptotic colors derived from their color profiles 
($(g-z)\sim 1$ mag) would locate them in the blue edge of the linear trend
of the CMR for their luminosities. This might be pointing towards dE/dIrr 
galaxies as previous stages of dE galaxies which define the blue border of the 
CMR, or as objects on their way towards the passive CMR. However, it is
necessary to spectroscopically confirm more low-mass cluster members before 
being able to propose a solid hypothesis about their origin on an
observational basis.

\acknowledgments
We thank the anonymous referee for her/his detailed revision and 
useful analysis of our paper which helped to greatly improve its content. 
We also thank Sergio Cellone for his help with the organization
of the images, color maps and profiles of the galaxies
in the webpage of our group, as well as for a fruitful discussion
of the manuscript. All of the data presented in this paper were 
obtained from the Mikulski Archive for Space Telescopes (MAST). STScI 
is operated by the Association of Universities for Research in Astronomy, 
Inc., under NASA contract NAS5-26555. Support for MAST for non-HST data is 
provided by the NASA Office of Space Science via grant NNX09AF08G and by 
other grants and contracts. This work was funded with grants from Agencia de
Promoci\'on Cient\'ifica y Tecnol\'ogica of Argentina (BID AR PICT 
2010-0410), Consejo Nacional de Investigaciones Cient\'ificas y T\'ecnicas 
de la Rep\'ublica Argentina (PIP-11220080100712, PIP-11220080101611) and 
Facultad de Ciencias Astron\'omicas y Geof\'isicas (Universidad Nacional de 
La Plata, Argentina).

\begin{deluxetable}{ccccccccc}
\tablenum{1}
\label{Table1}
\tablecolumns{8}
\tablewidth{0pc}
\tablecaption{Our photometry of the 100 galaxies included in the ACSVCS.}
\tablehead{
\colhead{} & \colhead{VCC} & \colhead{\it g} & \colhead{\it (g-z)} & \colhead{$r_{\rm{eff}_g}$} &  \colhead{$\langle \mu_{\rm{eff}} \rangle_{\Large g}$} & \colhead{$\epsilon$} & \colhead{$\theta$}\\ 
\colhead{} & \colhead{} & \colhead{\it (mag)} & \colhead{\it (mag)} & \colhead{\it (arcsec)} & \colhead{\it (mag arcsec$^{-2}$)} & \colhead{} & \colhead{\it (degrees)}}
\startdata
1   & 9     & 14.05  $\pm$ 0.01 &  1.07  $\pm$ 0.02 &  25.51  & 23.08	& 0.20&  50\\ 
2   & 21    & 14.83  $\pm$ 0.02 &  0.88  $\pm$ 0.03 &   9.02  & 21.61	& 0.30& -30\\ 
3   & 33    & 15.07  $\pm$ 0.02 &  1.13  $\pm$ 0.03 &   8.41  & 21.69 	& 0.20&  15\\ 
4   & 140   & 14.15  $\pm$ 0.02 &  1.20  $\pm$ 0.02 &   8.78  & 20.86 	& 0.10&  45\\ 
5   & 200   & 14.89  $\pm$ 0.01 &  1.17  $\pm$ 0.02 &  11.48  & 22.18	& 0.10&   9\\ 
6   & 230   & 15.60  $\pm$ 0.02 &  1.17  $\pm$ 0.03 &   8.55  & 22.26	& 0.10& -45\\ 
7   & 355   & 12.22  $\pm$ 0.01 &  1.50  $\pm$ 0.01 &   7.15  & 18.49	& 0.10&  45\\ 
8   & 369   & 11.93  $\pm$ 0.01 &  1.53  $\pm$ 0.01 & \nodata & \nodata	& 0.10&   9\\ 
9   & 437   & 14.17  $\pm$ 0.02 &  1.25  $\pm$ 0.02 &  17.42  & 22.37	& 0.30& -35\\ 
10  & 538   & 16.02  $\pm$ 0.02 &  1.15  $\pm$ 0.02 &   4.37  & 21.22	& 0.10&  -4\\ 
11  & 543   & 14.38  $\pm$ 0.01 &  1.20  $\pm$ 0.02 &  14.54  & 22.19	& 0.40&  45\\ 
12  & 571   & 15.13  $\pm$ 0.02 &  1.10  $\pm$ 0.03 &   9.45  & 22.01	& 0.40&  -4\\ 
13  & 575   & 13.91  $\pm$ 0.03 &  1.33  $\pm$ 0.04 &   5.48  & 19.59	& 0.10&  -4\\ 
14  & 654   & 12.08  $\pm$ 0.03 &  1.44  $\pm$ 0.03 & \nodata & \nodata & 0.30& -15\\ 
15  & 685   & 11.69  $\pm$ 0.05 &  1.49  $\pm$ 0.07 &  11.68  & 19.03	& 0.30&  90\\ 
16  & 698   & 13.25  $\pm$ 0.03 &  1.31  $\pm$ 0.04 &  13.14  & 20.84	& 0.30& -30\\ 
17  & 731   & 10.90  $\pm$ 0.01 &  1.54  $\pm$ 0.01 & \nodata & \nodata	& 0.30&  90\\ 
18  & 751   & 14.70  $\pm$ 0.02 &  1.27  $\pm$ 0.03 &   9.25  & 21.53	& 0.20&  -4\\ 
19  & 759   & 11.64  $\pm$ 0.02 &  1.54  $\pm$ 0.03 & \nodata & \nodata & 0.40& -25\\ 
20  & 763   & 10.49  $\pm$ 0.01 &  1.56  $\pm$ 0.01 & \nodata & \nodata	& 0.10&   0\\ 
21  & 778   & 12.33  $\pm$ 0.02 &  1.42  $\pm$ 0.02 &   8.40  & 18.94	& 0.30&  45\\  
22  & 784   & 12.28  $\pm$ 0.02 &  1.42  $\pm$ 0.02 &  11.97  & 19.67	& 0.30&  0 \\ 
23  & 798   & 10.58  $\pm$ 0.02 &  1.42  $\pm$ 0.02 & \nodata & \nodata	& 0.30&  90\\ 
24  & 828   & 12.78  $\pm$ 0.01 &  1.46  $\pm$ 0.02 &   9.38  & 19.64	& 0.30&  45\\ 
25  & 856   & 14.19  $\pm$ 0.02 &  1.18  $\pm$ 0.03 &  14.51  & 22.00	& 0.20&   9\\ 
26  & 881   & 10.61  $\pm$ 0.01 &  1.53  $\pm$ 0.01 & \nodata & \nodata	& 0.30& -20\\ 
27  & 944   & 11.97  $\pm$ 0.02 &  1.42  $\pm$ 0.03 & \nodata & \nodata	& 0.50& -65\\ 
28  & 1025  & 12.70  $\pm$ 0.01 &  1.41  $\pm$ 0.02 &   9.42  & 19.57	& 0.20&  40\\ 
29  & 1030  & 11.55  $\pm$ 0.02 &  1.48  $\pm$ 0.04 &  12.70  & 19.06	& 0.20&  59\\ 
30  & 1049  & 14.93  $\pm$ 0.02 &  1.08  $\pm$ 0.02 &   7.07  & 21.17	& 0.10& -45\\ 
31  & 1062  & 11.18  $\pm$ 0.02 &  1.50  $\pm$ 0.03 & \nodata & \nodata	& 0.50& -25\\ 
32  & 1075  & 14.95  $\pm$ 0.02 &  1.14  $\pm$ 0.03 &  15.52  & 22.90	& 0.10& -60\\ 
33  & 1087  & 13.97  $\pm$ 0.01 &  1.25  $\pm$ 0.01 &  16.69  & 22.08	& 0.30&   0\\ 
34  & 1125  & 13.02  $\pm$ 0.03 &  1.33  $\pm$ 0.05 & \nodata & \nodata	& 0.60& -80\\ 
35  & 1146  & 12.76  $\pm$ 0.02 &  1.34  $\pm$ 0.02 &  13.33  & 20.38	& 0.15&  45\\ 
36  & 1154  & 11.35  $\pm$ 0.01 &  1.59  $\pm$ 0.01 & \nodata & \nodata	& 0.20&   2\\ 
37  & 1178  & 13.14  $\pm$ 0.02 &  1.44  $\pm$ 0.03 &   5.82  & 18.96	& 0.15&  59\\ 
38  & 1185  & 15.56  $\pm$ 0.02 &  1.27  $\pm$ 0.02 &  14.13  & 23.31	& 0.10& -45\\ 
39  & 1192  & 14.90  $\pm$ 0.02 &  1.46  $\pm$ 0.03 &   4.73  & 20.26	& 0.05&   0\\ 
40  & 1199  & 16.44  $\pm$ 0.02 &  1.54  $\pm$ 0.02 &   2.09  & 20.03	& 0.10&  -4\\ 
41  & 1226  & 10.02  $\pm$ 0.01 &  1.57  $\pm$ 0.01 & \nodata & \nodata	& 0.25&  50\\ 
42  & 1231  & 11.16  $\pm$ 0.02 &  1.50  $\pm$ 0.02 & \nodata & \nodata	& 0.30& -15\\ 
43  & 1242  & 12.34  $\pm$ 0.03 &  1.39  $\pm$ 0.04 &  13.50  & 19.99	& 0.20&  80\\ 
44  & 1250  & 12.86  $\pm$ 0.02 &  1.29  $\pm$ 0.03 &  11.09  & 20.07	& 0.20&  85\\ 
45  & 1261  & 13.59  $\pm$ 0.01 &  1.19  $\pm$ 0.02 &  14.96  & 21.46	& 0.30&  30\\ 
46  & 1279  & 11.96  $\pm$ 0.01	&  1.44  $\pm$ 0.01 &  11.09  & 19.18	& 0.20&  45\\ 
47  & 1283  & 13.22  $\pm$ 0.02 &  1.42  $\pm$ 0.03 &  15.20  & 21.12	& 0.10&  40\\ 
48  & 1297  & 14.19  $\pm$ 0.02 &  1.52  $\pm$ 0.02 &   2.32  & 18.01	& 0.10&  45\\ 
49  & 1303  & 12.89  $\pm$ 0.02 &  1.41  $\pm$ 0.02 &  11.00  & 20.10	& 0.40& -55\\ 
50  & 1316  & 10.21  $\pm$ 0.01 &  1.59  $\pm$ 0.01 & \nodata & \nodata	& 0.15&  45\\ 
51  & 1321  & 12.77  $\pm$ 0.02 &  1.30  $\pm$ 0.02 &  14.95  & 20.64	& 0.20&  50\\ 
52  & 1327  & 13.12  $\pm$ 0.02 &  1.52  $\pm$ 0.03 &   7.71  & 19.55	& 0.20&  55\\ 
53  & 1355  & 14.69  $\pm$ 0.01 &  1.21  $\pm$ 0.03 &  20.63  & 23.26	& 0.10&  45\\ 
54  & 1407  & 15.04  $\pm$ 0.02 &  1.19  $\pm$ 0.01 &  10.32  & 22.10	& 0.10&  45\\ 
55  & 1422  & 13.73  $\pm$ 0.01 &  1.19  $\pm$ 0.01 &  17.44  & 21.93	& 0.20&  45\\ 
56  & 1431  & 14.35  $\pm$ 0.01 &  1.37  $\pm$ 0.01 &   9.25  & 21.17	& 0.10&   4\\ 
57  & 1440  & 14.79  $\pm$ 0.01 &  1.17  $\pm$ 0.02 &   6.78  & 20.94	& 0.10&  -9\\ 
58  & 1475  & 13.13  $\pm$ 0.02 &  1.29  $\pm$ 0.03 &   8.20  & 19.69	& 0.30&  50\\ 
59  & 1488  & 14.86  $\pm$ 0.02 &  0.91  $\pm$ 0.02 &   9.02  & 21.63	& 0.30& -40\\ 
60  & 1489  & 15.98  $\pm$ 0.02 &  1.06  $\pm$ 0.03 &  10.00  & 22.97	& 0.30& -45\\ 
61  & 1499  & 15.01  $\pm$ 0.03 &  0.76  $\pm$ 0.04 &   5.39  & 20.66	& 0.10& -45\\ 
62  & 1512  & 15.80  $\pm$ 0.02 &  1.16  $\pm$ 0.03 &   8.55  & 22.45	& 0.20& -45\\ 
63  & 1528  & 14.49  $\pm$ 0.01 &  1.23  $\pm$ 0.01 &   8.83  & 21.21	& 0.10&   0\\ 
64  & 1535  & 10.69  $\pm$ 0.02 &  1.56  $\pm$ 0.03 & \nodata & \nodata & 0.50&   4\\ 
65  & 1537  & 12.56  $\pm$ 0.02 &  1.43  $\pm$ 0.03 &   7.99  & 19.07	& 0.30&  65\\ 
66  & 1539  & 15.86  $\pm$ 0.02 &  1.22  $\pm$ 0.02 &  14.27  & 23.63	& 0.10&  45\\ 
67  & 1545  & 14.90  $\pm$ 0.03 &  1.26  $\pm$ 0.04 &   9.31  & 21.74	& 0.30& -15\\ 
68  & 1619  & 12.24  $\pm$ 0.02 &  1.39  $\pm$ 0.02 &  10.63  & 19.37	& 0.60&  65\\ 
69  & 1627  & 15.26  $\pm$ 0.01 &  1.41  $\pm$ 0.01 &   3.44  & 19.93	& 0.10&   9\\ 
70  & 1630  & 12.64  $\pm$ 0.01 &  1.50  $\pm$ 0.02 &  11.35  & 19.91	& 0.20& -18\\ 
71  & 1632  & 10.95  $\pm$ 0.01 &  1.59  $\pm$ 0.01 & \nodata & \nodata & 0.10&  9 \\ 
72  & 1661  & 15.89  $\pm$ 0.02 &  1.18  $\pm$ 0.03 &  16.08  & 23.92	& 0.10&  9 \\ 
73  & 1664  & 11.91  $\pm$ 0.02 &  1.48  $\pm$ 0.03 & \nodata & \nodata	& 0.40& -55\\ 
74  & 1692  & 11.61  $\pm$ 0.02 &  1.48  $\pm$ 0.03 & \nodata & \nodata	& 0.60&  50\\ 
75  & 1695  & 14.38  $\pm$ 0.02 &  1.12  $\pm$ 0.02 &  19.52  & 22.83	& 0.10&  -4\\ 
76  & 1720  & 12.17  $\pm$ 0.02 &  1.42  $\pm$ 0.03 &  17.42  & 20.37	& 0.30&  80\\ 
77  & 1743  & 15.53  $\pm$ 0.03 &  1.05  $\pm$ 0.05 &  10.62  & 22.65	& 0.30&  35\\ 
78  & 1779  & 14.82  $\pm$ 0.03 &  1.00  $\pm$ 0.04 &  10.88  & 22.00	& 0.50& -45\\ 
79  & 1826  & 15.55  $\pm$ 0.01 &  1.13  $\pm$ 0.02 &   6.27  & 21.53	& 0.30&  35\\ 
80  & 1828  & 15.23  $\pm$ 0.02 &  1.15  $\pm$ 0.03 &  12.80  & 22.76	& 0.20&  45\\ 
81  & 1833  & 14.53  $\pm$ 0.02 &  1.21  $\pm$ 0.03 &   7.07  & 20.77	& 0.10&   0\\ 
82  & 1857  & 15.01  $\pm$ 0.02 &  0.94  $\pm$ 0.02 &  19.17  & 23.42	& 0.40&   0\\ 
83  & 1861  & 14.20  $\pm$ 0.01 &  1.28  $\pm$ 0.02 &  15.94  & 22.21	& 0.10&  45\\ 
84  & 1871  & 14.19  $\pm$ 0.02 &  1.42  $\pm$ 0.02 &   6.41  & 20.22	& 0.10&   4\\ 
85  & 1883  & 12.07  $\pm$ 0.01 &  1.33  $\pm$ 0.02 &  14.35  & 19.85	& 0.30&  80\\ 
86  & 1886  & 15.37  $\pm$ 0.01 &  1.01  $\pm$ 0.02 &  12.26  & 22.81	& 0.40&  65\\ 
87  & 1895  & 14.99  $\pm$ 0.02 &  1.08  $\pm$ 0.03 &   9.29  & 21.83	& 0.50& -60\\ 
88  & 1903  & 10.94  $\pm$ 0.01 &  1.55  $\pm$ 0.02 & \nodata & \nodata	& 0.30&  45\\ 
89  & 1910  & 14.24  $\pm$ 0.01 &  1.36  $\pm$ 0.01 &  11.25  & 21.49	& 0.10&  45\\ 
90  & 1913  & 13.05  $\pm$ 0.03 &  1.33  $\pm$ 0.04 &  13.17  & 20.65	& 0.50&  45\\ 
91  & 1938  & 11.95  $\pm$ 0.03 &  1.42  $\pm$ 0.04 &   9.43  & 18.82	& 0.30&  45\\ 
92  & 1948  & 15.47  $\pm$ 0.02 &  1.02  $\pm$ 0.03 &  10.82  & 22.64	& 0.40& -20\\ 
93  & 1978  & 10.25  $\pm$ 0.01 &  1.63  $\pm$ 0.01 & \nodata & \nodata & 0.20&  -9\\ 
94  & 1993  & 15.48  $\pm$ 0.02 &  1.19  $\pm$ 0.02 &   9.44  & 22.35	& 0.10&  4 \\ 
95  & 2000  & 11.94  $\pm$ 0.03 &  1.43  $\pm$ 0.04 &   8.55  & 18.60	& 0.20&  -9\\ 
96  & 2019  & 14.43  $\pm$ 0.01 &  1.15  $\pm$ 0.02 &  14.55  & 22.24	& 0.20& -80\\ 
97  & 2048  & 14.00  $\pm$ 0.01 &  1.18  $\pm$ 0.02 &  10.08  & 21.01	& 0.50&  65\\ 
98  & 2050  & 15.22  $\pm$ 0.03 &  1.16  $\pm$ 0.04 &  10.52  & 22.33	& 0.20&  85\\ 
99  & 2092  & 11.39  $\pm$ 0.02 &  1.51  $\pm$ 0.03 & \nodata & \nodata & 0.40& -80\\ 
100 & 2095  & 11.69  $\pm$ 0.04 &  1.39  $\pm$ 0.06 & \nodata & \nodata	& 0.40& -15\\ 
\enddata
\tablecomments{Col. (2) shows the VCC designation of the galaxies. Col. (3) 
shows not extinction corrected $g$ magnitudes. Col. (4) lists not reddening 
corrected $(g-z)$ colors. Col. (5) shows $r_{\rm eff}$ obtained in this work 
in the $g$ band for galaxies that are fully contained in the ACS FOV. Col. (6) 
lists the corresponding $\langle \mu_{\rm eff} \rangle$ in the $g$ band. 
Cols. (7) and (8) show the fixed ellipticity and position angle used to 
obtain brightness and color profiles from which integrated magnitudes and 
colors were calculated.} 
\end{deluxetable}

\begin{deluxetable}{lccccc}
\tablenum{2}
\label{ajustes}
\tablecolumns{6}
\tablewidth{0pc}
\tablecaption{Results of least-square fits 
performed to the linear region of the CMR of early-type
galaxies in the Virgo cluster} 
\tablehead{
\colhead{Sample} & \colhead{Data} & \colhead{$a$} & \colhead{$b$} & \colhead{$r$} & \colhead{$\sigma_{\small(g-z)_0}$}\\
\\
\hline
\\
\multicolumn{6}{c}{\citet{Ferrarese06}}}
\startdata
Whole trend    		& 74 & 1.28 $\pm$ 0.010 & -0.097 $\pm$ 0.008 & -0.84 & 0.06 \\ 
Bright region  		& 35 & 1.31 $\pm$ 0.026 & -0.069 $\pm$ 0.020 & -0.58 & 0.07 \\
Faint region   		& 39 & 1.34 $\pm$ 0.021 & -0.163 $\pm$ 0.017 & -0.47 & 0.10 \\
Reduced sample 		& 61 & 1.28 $\pm$ 0.012 & -0.098 $\pm$ 0.011 & -0.79 & 0.06 \\
Bright Reduced sample 	& 25 & 1.34 $\pm$ 0.035 & -0.034 $\pm$ 0.036 & -0.46 & 0.09 \\
Faint Reduced sample 	& 36 & 1.23 $\pm$ 0.025 & -0.087 $\pm$ 0.017 & -0.49 & 0.04 \\
\cutinhead{Our work}
Whole trend    		& 64 & 1.27 $\pm$ 0.011 & -0.089 $\pm$ 0.009 & -0.83 & 0.07 \\
Bright region  		& 26 & 1.28 $\pm$ 0.025 & -0.094 $\pm$ 0.029 & -0.51 & 0.06 \\
Faint region   		& 38 & 1.21 $\pm$ 0.034 & -0.046 $\pm$ 0.023 & -0.35 & 0.08 \\
Reduced sample 		& 61 & 1.27 $\pm$ 0.012 & -0.089 $\pm$ 0.009 & -0.83 & 0.07 \\
Bright Reduced sample 	& 24 & 1.28 $\pm$ 0.025 & -0.101 $\pm$ 0.030 & -0.55 & 0.06 \\
Faint Reduced sample 	& 37 & 1.21 $\pm$ 0.034 & -0.047 $\pm$ 0.024 & -0.38 & 0.07 \\
\cutinhead{\citet{Chen10}}
Whole trend             & 79 & 1.27 $\pm$ 0.012 & -0.078 $\pm$ 0.008 & -0.81 & 0.07 \\
Bright region           & 42 & 1.28 $\pm$ 0.024 & -0.070 $\pm$ 0.015 & -0.64 & 0.05 \\
Faint region            & 37 & 1.19 $\pm$ 0.033 & -0.032 $\pm$ 0.026 & -0.31 & 0.13 \\
Reduced sample          & 61 & 1.27 $\pm$ 0.011 & -0.082 $\pm$ 0.010 & -0.74 & 0.07 \\
Bright Reduced sample 	& 25 & 1.28 $\pm$ 0.033 & -0.066 $\pm$ 0.032 & -0.52 & 0.08 \\
Faint Reduced sample 	& 36 & 1.19 $\pm$ 0.034 & -0.033 $\pm$ 0.026 & -0.31 & 0.06 \\
\enddata
\tablecomments{The fitted function is of the form  $(g-z)_0=a+b~(M_{g_0}+18)$.
Col. (1) indicates the different samples considered to perform
the fits. Col. (2) gives the number of data points. Col. (3) and (4) list 
the slopes and zero-points of the reddening and extinction corrected mean 
linear CMRs. Col. (5) shows the Pearson correlation coefficient, $r$, of the 
different fits. Col. (6) presents the intrinsic color scatter which was 
obtained as in \citet{SC12}. The adopted limiting magnitude to separate the 
bright and faint regions of the linear trends is that of the observed 
intermediate gap ($M_{g_0} \sim -17.5$ mag).}
\end{deluxetable}
 
\appendix

\section{Our values compared with those of F06}
\label{valores}

\begin{figure*}
\center
\includegraphics[scale=0.85]{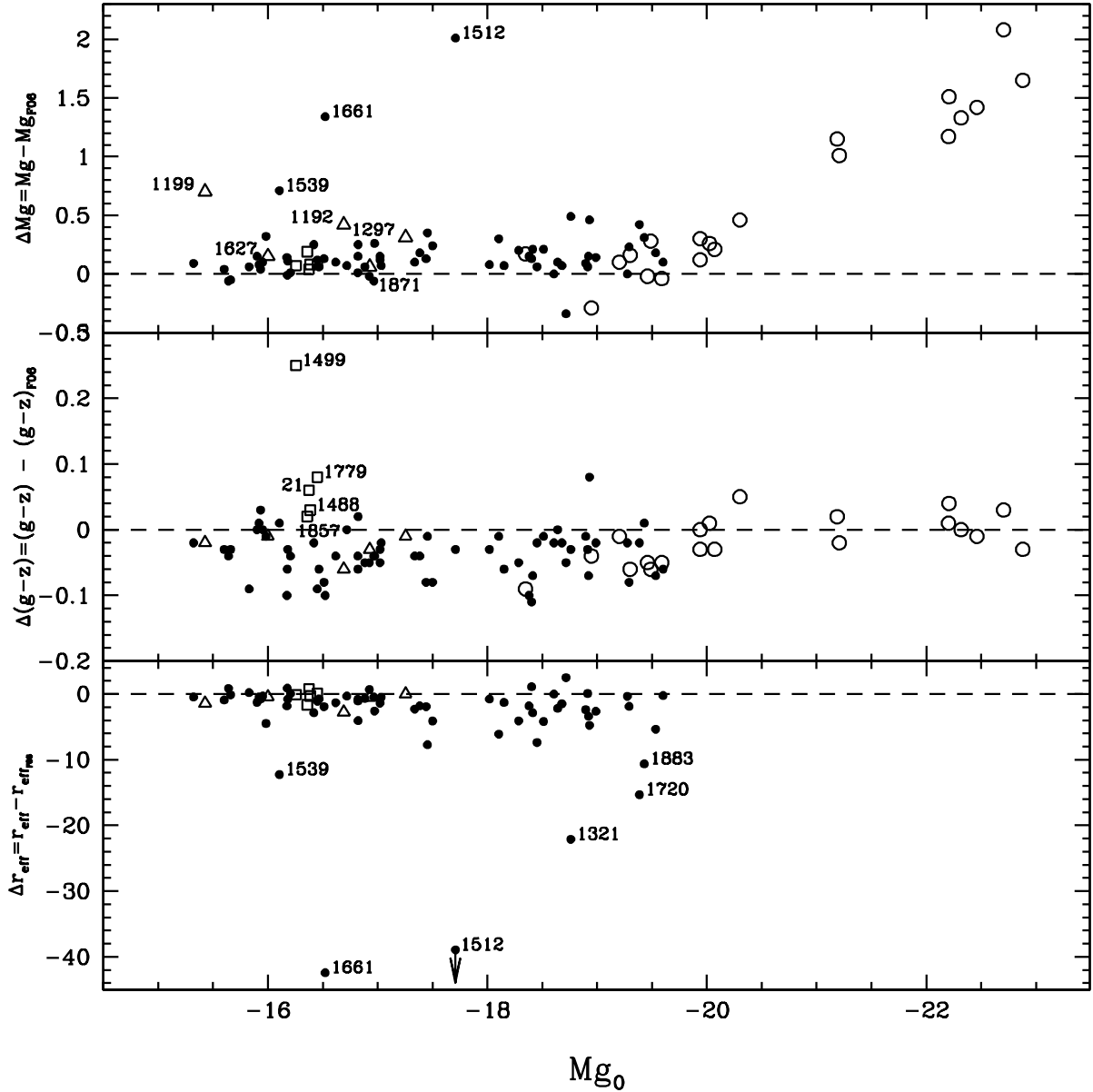}
\caption{We show the differences between our $g$ magnitudes {\it(top)}, 
$(g-z)$ colors {\it (center)} and $r_{\rm eff}$ values in arcsec 
{\it (bottom)}, and those reported by F06 for the galaxies of the ACSVCS. We 
depict with open circles incomplete galaxies. For clarity, in the top plot we 
only label the cE galaxies {\it (open triangles)} and in the centered one, the 
blue galaxies {\it (open squares)}. In all the plots we also label the most 
discrepant galaxies that are complete within the ACS frames.}
\label{delta}
\end{figure*}

In Figure\,\ref{delta} we present three plots that show, as a function of 
luminosity, the differences between our $g$ magnitudes, $(g-z)$ colors and 
$r_{\rm eff}$ values, and those found by F06 ($\Delta M_g=M_g - M_{g_{F06}}$, 
$\Delta (g-z)=(g-z)-(g-z)_{F06}$ and 
$\Delta r_{\rm eff}=r_{\rm eff}-r_{{\rm eff}_{F06}}$, respectively). We can 
see that our magnitudes tend to be fainter, our colors bluer and our 
$r_{\rm eff}$ values smaller than those of F06. This is consistent with the 
fact that we have isophotal limited brightnesses while those of F06 are 
obtained from the integration to infinity of theoretic brightness profiles. In 
addition, it is expected that our colors are bluer than those of F06 as 
early-type galaxies tend to be redder in their central regions, becoming bluer 
outwards. Our colors are calculated as the difference between the integrated 
magnitudes obtained from the integration of the observed brightness profiles.
F06 measured their colors between $1''$ and 1 $r_{\rm eff}$.

In particular, for complete galaxies, we obtain 
$\langle \Delta M_g \rangle=0.13$ mag excluding the four more 
separated galaxies, 
$\langle \Delta (g-z) \rangle=-0.03$ mag excluding the two
more separated objects, and $\langle \Delta r_{\rm eff} \rangle=-1.53''$ 
excluding the six more separated values.

The most important deviation in color is shown by VCC\,1499 
($\langle \Delta (g-z) \rangle=0.25$ mag). VCC\,1499 display a strong 
gradient within 1 $r_{\rm eff}$. Therefore, this discrepancy probably arises 
because F06 exclude the external (and reddest) regions of the galaxy to obtain 
its color.  

The largest deviations both in luminosity ($\Delta M_g=2.01$ mag) 
and $r_{\rm eff}$ ($\Delta r_{\rm eff}=-488.95''$) correspond to 
VCC\,1512, followed by those of VCC\,1661 ($\Delta M_g=1.34$ mag and 
$\Delta r_{\rm eff}=-42.42''$). The case of VCC\,1512 was 
already noticed in Section\,\ref{CMR_F06} and is related to unreliable 
quantities as reported by F06. In the case of VCC\,1661, the 
$g$ brightness profile reaches $\mu_{g_0}\approx27$ mag arcsec$^{-2}$ at 
$r\approx40''$. Therefore, the value of $r_{\rm eff}=58.50''$ 
reported by F06 is clearly overestimated, giving also an overestimated 
luminosity. VCC\,1539 presents a similar situation though not so severe, 
as F06 reported $r_{\rm eff}=26.53''$, but its brightness profile reaches 
$\mu_{g_0}\approx27$ mag arcsec$^{-2}$ at $r\approx30''$.

VCC\,1321, VCC\,1720 and VCC\,1883 show significant differences in 
$r_{\rm eff}$ but not in integrated magnitudes. This situation clearly 
points towards a discrepancy arising because of the different methods 
used to obtain $r_{\rm eff}$. It is difficult to say if the fitting 
method overestimates the values, or the isophotal limited analysis 
underestimate them. This would deserve an analysis that it is 
beyond the scope of this paper. However, 
it can be noticed that in the case of VCC\,1883, we have obtained a  
brightness profile that would prevent a good fit even considering two 
components.

We also point out that when we measure integrated colors in the same way as 
F06 do (that is, integrating the brightness profiles between $1''$ and 1 
$r_{\rm eff}$), we obtain 
$\langle (g-z)_{reff} - (g-z)_{F06}\rangle=0.009\pm0.03$ mag.

\end{document}